\documentclass[conference,doublecolumn]{IEEEtran} % IEEEtran

\usepackage{comment}
\usepackage{verbatim}
\usepackage{amsthm}
\usepackage{amssymb}
\usepackage{xcolor}
\usepackage{float}
\usepackage{amsmath}
\usepackage{subcaption}
\usepackage{graphicx}% Include figure files
\usepackage{dcolumn}% Align table columns on decimal point
\usepackage{bm}% bold math
\usepackage[colorlinks=true, allcolors=blue]{hyperref}
\usepackage[mathlines]{lineno}% Enable numbering of 

\usepackage[utf8]{inputenc}
\usepackage{booktabs}
\usepackage{tabularx}
\usepackage{xcolor}
\usepackage{colortbl}

\definecolor{hispasatblue}{rgb}{0.098,0.2509, 0.4470}

\begin{document}
\bstctlcite{BSTcontrol}

\title{Atmospheric channel model for QKD over satellite megaconstellations}

\author{
\IEEEauthorblockN{Sergio Gándara Gándara, Javier Vecino Peñas, Ana Fernández-Vilas, \\
Manuel Fernández-Veiga, Rebeca P. Díaz Redondo}
\IEEEauthorblockA{
atlanTTic Research Centre, Universidade de Vigo \\ \{sergio, javier.vecino, avilas, mveiga, rebeca\}@det.uvigo.es
}
\thanks{Corresp. author: javier.vecino@det.uvigo.es}
}

\maketitle

\begin{abstract}
Atmospheric turbulence and channel loss represent the primary limiting factors in 
ground-to-satellite quantum communication systems, yet their joint impact remains 
insufficiently characterized across diverse operational conditions. In this work, 
we develop a comprehensive and open atmospheric channel model for quantum optical links along 
ground-to-satellite paths, integrating altitude-dependent molecular and aerosol absorptions. 
The model is validated with available data from missions such as Micius and Hispasat, 
including a secondary viability assessment for a hypothetical Starlink-based quantum
communication network, where we evaluate the predicted channel behavior in terms of 
single-photon transmittance and power balance, alongside the channel capacity of the 
link for uplink and downlink exclusively. Our comparative analysis demonstrates the ability of our channel model to predict real data under simple atmospheric behavior assumptions, which offer practical guidelines for the design and optimization of future quantum satellite missions.
\end{abstract}
\begin{IEEEkeywords}
 Quantum key Distribution (QKD), Satellite QKD, Atmospheric Channel Model, 
Transmissivity,  Channel Capacity
\end{IEEEkeywords}

\section{Introduction} 
\label{sec:introduction}
The growing interest in quantum communication networks has driven significant research 
efforts toward establishing reliable ground-to-satellite quantum links. Unlike classical 
optical communication, quantum key distribution (QKD) and other quantum protocols are 
highly sensitive to photon loss, noise and decoherence introduced by the transmission
medium. The atmospheric channel, in particular, imposes a fundamental challenge, as 
turbulence, absorption and scattering translates into severe constraints on the fidelity 
and efficiency of quantum states transmission.

Ground-to-satellite quantum links have emerged as a promising approach to extend quantum 
networks over global distances, bypassing the exponential loss limitations of fiber-based
infrastructures. Several milestone experiments~\cite{RevModPhys.94.035001}, most notably 
those carried out with the Micius satellite, have demonstrated the feasibility of 
satellite-based QKD, entanglement distribution and quantum teleportation over distances 
exceeding a thousand km. However, the performance of such systems depends critically on 
the atmospheric conditions encountered along the optical path, which vary with altitude, 
geographic location, season and time of day.

Accurate modeling of the atmospheric channel is therefore essential for the design, 
optimization and performance prediction of quantum communication systems operating 
along ground-to-satellite links. Key physical phenomena that must be accounted for 
include beam wandering and broadening induced by refractive index fluctuations, 
transmittance fluctuations, and altitude-dependent absorption due to molecular 
and aerosol species. Concepts borrowed from thermodynamics and optics, such as 
density-dependent refractive indices and Rayleigh scattering will provide the 
mathematical description to tackle these physical phenomena.

In this work, we present a comprehensive atmospheric channel model tailored 
to ground-to-satellite quantum optical links. The model integrates turbulence-induce 
index variations, beam propagation effects, and wavelength-dependent transmittance 
profiles derived from standard atmospheric databases. We evaluate the impact of these 
factors on key classical and quantum communication metrics, such as the single-photon
transmission probability and losses, power balance and channel capacity. Furthermore,
we benchmark our model against data extracted from real-world quantum satellite 
missions such as Micius and Hispasat. This comparative analysis allows us to identify 
the strengths and limitations of this approach, prove the feasibility of space Quantum Key 
Distribution and to explore future quantum satellite missions.

The remainder of the paper is organized as follows. We briefly review prior works on
the feasibility of satellite QKD in Section~\ref{sec:related-work}. Our physical modeling
of the atmospheric optical channel is presented in Section~\ref{sec:model}, and the 
technical specifications relevant for QKD satellite missions are thoroughly discussed in
Section~\ref{sec:specifications}. The validation of our model under realistic parameterization
and a comparison with experimental values is the focus of Section~\ref{sec:results}. Finally, Section~\ref{sec:conclusions} summarizes the contributions of this work.

\section{Related Work}
\label{sec:related-work}

The feasibility of satellite-based quantum communication was first theoretically 
analyzed in~\cite{Rarity_Tapster_Gorman_Knight_2002}, who identified the low photon 
loss of free-space links at high altitudes as a key advantage over fiber-based 
networks. Subsequent important works carried out in~\cite{Aspelmeyer_2003_satellite_quantum} 
and~\cite{Pfennigbauer2005} laid the groundwork for satellite QKD, addressing 
photon loss budgets and link efficiency under realistic atmospheric conditions.

On the experimental front, the launch of the Micius quantum science satellite in
2016 marked a turning point in the field. \cite{Liao_2018_satellite_quantum_network}
demonstrated intercontinental QKD over a distance exceeding 7,600 km using a 
trusted-node architecture, while~\cite{Yin_2017_satellite_entanglement} reported 
the distribution of entangled photon pairs between ground stations separated by 
over 1,200 km. These experiments confirmed the practical viability of 
satellite-mediated quantum links and established key benchmarks for atmospheric 
channel loss. The atmospheric channel has been extensively studied in the physical 
domain. The Kolmogorov turbulence theory~\cite{Kolmogorov_1941_turbulence} and 
the Hufnagel-Valley refractive index structure parameter 
profile~\cite{Hufnagel_Stanley_1964_MTF_turbulence} remain the dominant analytical 
frameworks for characterizing beam propagation through turbulent media. More sophisticated models that study the small-scale fading and interference statistics of a signal include generalizations of other well-known distributions such as Gaussian \cite{Vaezi2016Simplified} or Rayleigh distributions, called the $\lambda-\kappa-\mu$ distributions \cite{Wang2024LambdaKappaMu,Liu2025DoubleLambdaKappaMu,Wang2026LambdaKappaMuShadowed}, as well as other generalized distributions like $\eta-\mu$ \cite{Nallagonda2021Generalized}, $\kappa-\mu$ \cite{Porto2016PhaseStatistics}, $\eta-\kappa$ \cite{Nadarajah2006EtaKappa} and $\alpha-\eta-\mu/\text{IGA}$ \cite{Xu2026AlphaEtaMuIGA} can accurately predict signal fading by properly tuning the different model parameters.

Physics-based models such as ITU-R recommended models (for instance: ITU-R P.676-13 \cite{ITUR67613}) or geometrical optics models like the Yudachev-Pinhasi model \cite{Yudachev2026Integrative}, although rigorous and accurate, rely on large amount of atmospheric data and variables, hindering their code implementation.

Wavelength-dependent atmospheric transmittance has been modeled using radiative 
transfer codes such as MODTRAN~\cite{Berk_2006_MODTRAN5} and 
LOWTRAN~\cite{Kneizys_1988_LOWTRAN7}, which account for molecular absorption 
by water vapor, oxygen, and carbon dioxide, as well as aerosol scattering.

More recent efforts have focused on the comparison and validation of atmospheric
models under quantum communication-specific figures of merit. For instance, 
\cite{Dequal_2021_CVQKD_satellite} analyzed the performance of a ground-to-satellite 
QKD system under varying atmospheric conditions. Similarly, \cite{Sidhu_2021_space_quantum} 
conducted a systematic study of the parameter space governing satellite QKD performance,
identifying elevation angle, ground station aperture, and turbulence strength 
as the dominant factors.

Our model, although less rigorous than the models found in the literature, conforms a simple and straightforward method to obtain the atmospheric refractive index and probability of transmission, that is, a compromise between accuracy and simplicity, focusing on the latter.

\section{Theoretical Framework and Model Description} 
\label{sec:model}

\begin{figure}[t]
    \centering
    \includegraphics[width=0.43\textwidth]{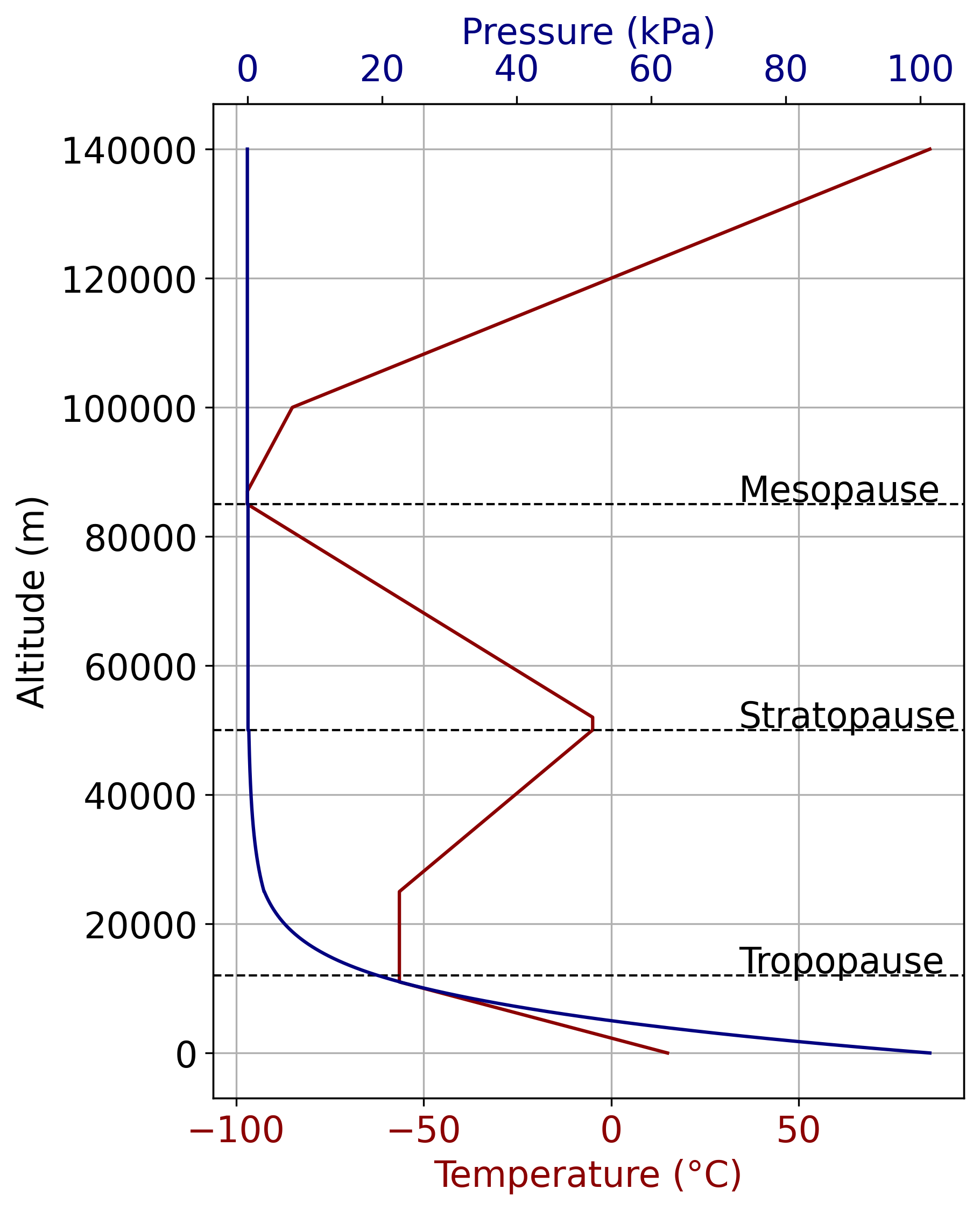}
    \caption{Pressure and Temperature profiles against altitude $z$ 
    (expressed in m).}
    \label{fig:PT}
\end{figure}

\subsection{Composition of the atmosphere}
The main obstacle between Earth's surface and satellites is the atmosphere. 
The atmosphere is a gas ensemble that surrounds the Earth composed of Nitrogen $(78\%)$, 
Oxygen $(21\%)$ and many other gasses such as Argon, Carbon dioxide, Ozone,
Water vapor and Helium. We can better describe the atmosphere by identifying five 
different layers: Troposphere, Stratosphere, Mesosphere, Thermosphere and Exosphere.
1) \textbf{Troposphere}: the lowest layer of the atmosphere and the place where 
the Earth's weather takes place. It extends to $\sim10$ km of altitude and the 
temperature decreases the deeper we reach. 2) \textbf{Stratosphere}: sits just 
above the Troposphere. It extends from the previous layer to around $50$ km and is 
the home of the Ozone layer. Due to the absorption of UV rays, 
the temperature increases with height. 3) \textbf{Mesosphere}: third 
highest atmospheric layer reaching a height of $80$ km. Some nitrogen-based molecules 
get weakly-ionized as a result of the presence of the Sun, causing a de-ionization 
at night. The temperature decreases with height. 4) \textbf{Thermosphere}:  
characterized by an increase of the temperature with height and a high density 
of ionized gases. It extends to $\sim 1000$ km. The line of $100$ km of 
height is named Karman's line and marks the starting point of the space. 
5) \textbf{Exosphere}: the furthest atmospheric layer. It is comprised of extremely 
low-pressure gasses such as Hydrogen and Helium.

We need to resort to the refractive index of each layer to study how light interacts 
with the atmospheric layers and how much light gets absorbed by each layer, but first 
we need to know the pressure profile $P(z)$ and temperature profile $T(z)$ with respect 
to altitude $z$. There are many low-altitude atmospheric
models~\cite{US_Standard_Atmosphere_1976, ICAO_1993_ISA} that model pressure and 
temperature as piecewise functions, so the natural step is, starting from these models, 
to be able to model $P(z)$ and $T(z)$ as piecewise functions that properly describe the measured 
pressures and temperatures. We sampled the real, measured values from the National Oceanic and Atmospheric 
Administration (NOAA) and approximated the values to follow simple mathematical relations. For the temperature profile $T(z)$, 
we get 
\begin{equation}
T(z) = 
\begin{cases} 
      15.04 - 0.0065\,z & z\leq 11 \,\text{km} \\
      -56.46 & 11< z\leq 25 \,\text{km}\\
      -107.91 + 0.0020584\, z & 25< z\leq 50 \,\text{km}\\
      -4.91 & 50 < z \leq 52 \,\text{km}\\
      140.17 - 0.00279\, z & 52< z\leq 85 \,\text{km}\\
      -97.0 & 85< z\leq 87 \,\text{km}\\
      -177.31 + 0.0009231\, z & 87< z\leq 100 \,\text{km}\\
      -510 + 0.00425\, z & z > 100 \,\text{km}\\
   \end{cases}
\end{equation}
 and, for the pressure profile  $P(z)$, we arrive at
\begin{equation}
P(z) = 
\begin{cases} 
      101.3\cdot \left( \dfrac{T(z)+273.1}{288.08}\right)^{5.256} & z\leq 11 \,\text{km} \\
      22.65 \cdot \exp(1.73 - 0.000157 \cdot z) & 11< z\leq 25 \,\text{km}\\
      2.488 \cdot \left( \dfrac{T(z)+273.1}{216.6}\right)^{-11.388} & 25< z\leq 50 \,\text{km}\\
      \sim 0 & z > 50 \,\text{km}
   \end{cases}
\end{equation}
where $T(z)$ is expressed in degrees Celsius. 

Figure~\ref{fig:PT} plots these functions, where we can clearly 
see the behavior  of pressure and temperature in the different atmospheric layers. 
From this, we  may compute now the densities of the gasses in the atmosphere. 
Assuming that  the atmospheric gasses behave as ideal gasses, then it is straightforward 
to  check that their density can be computed as~\cite{CengelBolesThermo}
\begin{equation}
   \label{eq:rho}
\rho_i (z) = \dfrac{P_i\,M_i}{R\, T} = 0.1203\cdot\chi_i \cdot P\, M_i/T,
\end{equation}
where $M_i$ stands for the molecular mass of each specie, $\chi_r$ stands for the molar fraction of the gasses. 
For water vapor, we have a similar expression \cite{Voronin_Zheltikov_Sellmeier_air}
\begin{equation}\label{eq:rhowv}
    \rho_{\text{H}_2\text{O}}(z) =0.1203 \cdot h P_s \,M_{\text{H}_2\text{O}}/T,
\end{equation}
where $h$ measures humidity and $P_s \equiv P_c \cdot 
\exp{\left[\tau\left(\alpha_1 \theta + \alpha_2\theta^{1.5}  + \alpha_3 \theta^{3} 
+ \alpha_4 \theta^{3.5} + \alpha_5 \theta^{4}  + \alpha_6 \theta^{7.5} \right)  \right]}$, 
with $\tau \equiv T_c/T$, $\theta \equiv 1- T/T_c$ and $T_c, \, P_c$ the temperature 
and pressure at the critical point of  $\text{H}_2\text{O}$.

%\begin{figure}[t]
%    \centering
%    \includegraphics[width=0.6\linewidth]{Figures/densities.png}
%    \caption{Densities of the different atmospheric against altitude.}
%    \label{fig:rho}
%\end{figure}

Since pressure $P(z)$ behaves as a decaying exponential, then the various densities must also obey an exponential law.
with  altitude due to the contribution of $P(z)$ on~\eqref{eq:rho}--\eqref{eq:rhowv}. 
The densities computed will allow us to define how the atmosphere absorbs light
on the following sections.

\subsection{Atmospheric refractive index}

It is known that, for a gas mixture, its refraction index must depend on the 
densities of the gasses that make up the total gas and on the wavelength of the 
incident light. As we previously noticed, the densities of the atmospheric gasses 
inherit the exponential behavior of atmospheric pressure, allowing us to describe 
the atmospheric refractive index as a function dependent on $P(z)$. We also know that, 
for high altitude $z$ ($P(z) \rightarrow0$), the atmospheric refractive index must 
tend to the value of 1 since, for high altitudes, we can approximate the atmosphere 
to the spatial vacuum. Taking these considerations into account, we propose the 
following expression for the refractive index
\begin{equation}\label{eq:index}
    n(P, \lambda) \approx 1 + \left[n_0(\lambda) - 1\right]\cdot P(z) /P_0,
\end{equation}
where $P$ is the atmospheric pressure dependent on altitude, $P_0 \equiv 1\, 
\text{atm}$ and $n_0(\lambda) = 1 + \Lambda_0 /\lambda$. The parameter 
$\Lambda_0$ encodes the strength the wavelength's contribution has on the total 
refractive index, its value will be set to $\Lambda_0 = 15$ nm since it reproduces 
the correct earth surface's refractive index for the wavelength range close to 
$\lambda = 1000 \,\text{nm}$. 

\begin{figure}[t]
    \centering
    \includegraphics[width=0.5\textwidth]{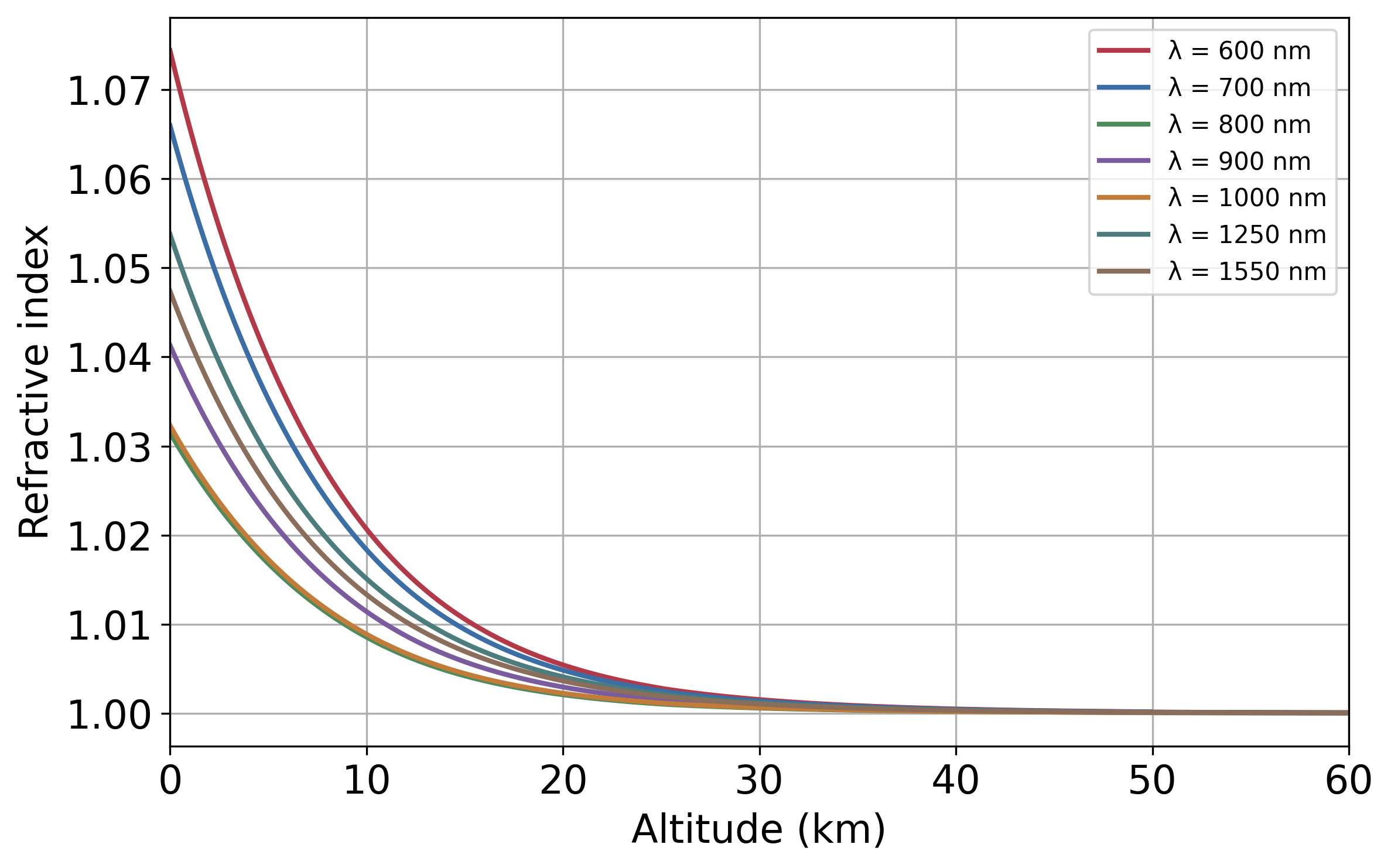}
    \caption{Refractive indices against altitude $z$ for various wavelengths.}
    \label{fig:index}
\end{figure}

Earth's weather takes place in the lowest layers of the atmosphere, we could 
incorporate a small, random coefficient $\delta_n$ as a correction
$n_R(P,\lambda) \equiv n(P,z) + \delta_r$, $\delta_r\ll n$.
The parameter $\delta_r$ will exponentially decrease with altitude, since in 
the upper layers of the atmosphere there are no weather effects. Figure~\ref{fig:index}
shows the variation for wavelength $\lambda = 1000 \, \text{nm}$.

\subsection{Propagation, scattering and absorption}
\begin{figure}[t]
    \centering
    \includegraphics[width=0.5\textwidth]{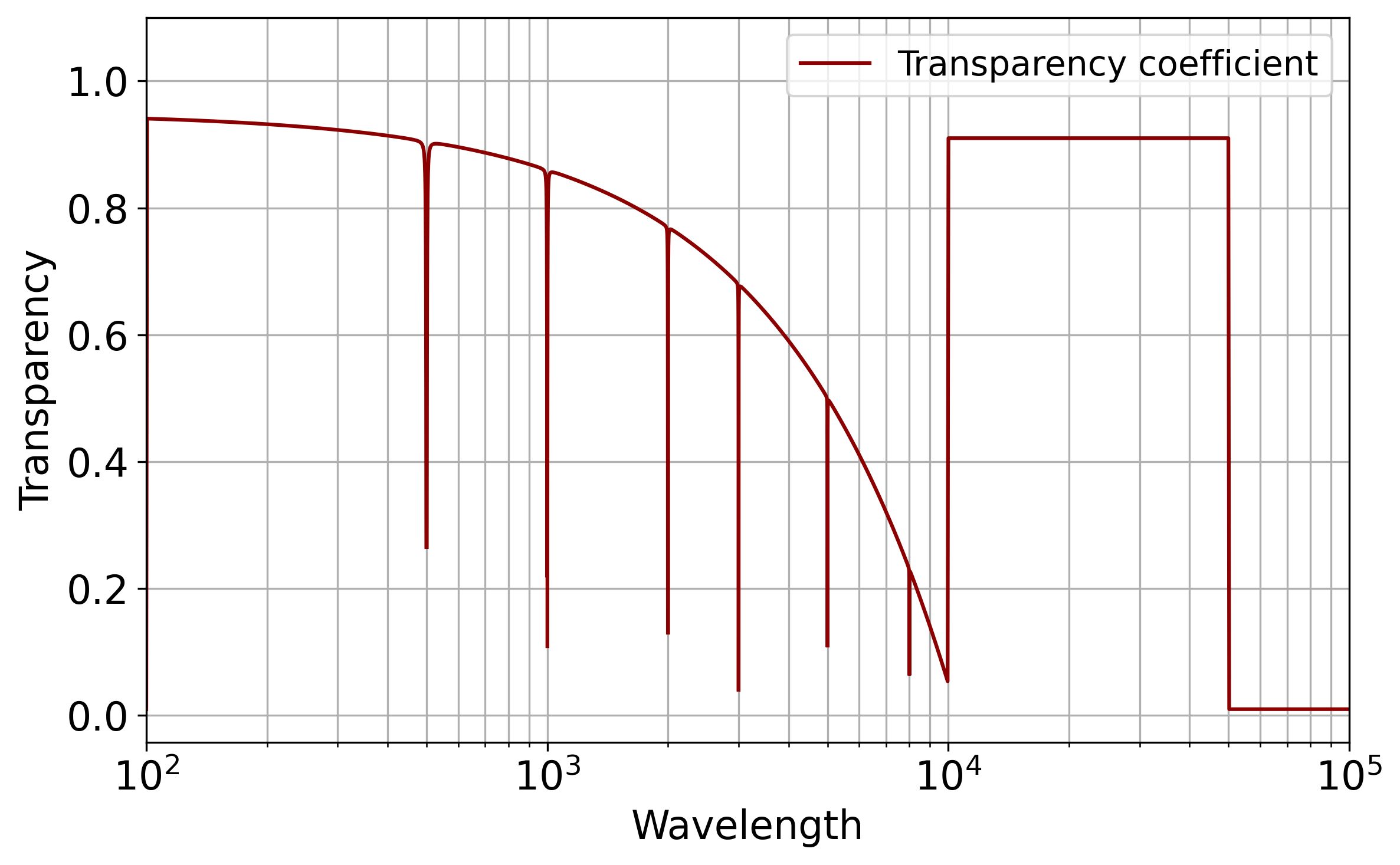}
    \caption{\textit{Transparency} against wavelength from 100 nm to 100 \text{$\mu$}m (log scale).}
    \label{fig:transp}
\end{figure}

When a wave propagates through a medium, its intensity decreases. 
Theoretically, we can explain this phenomenon if we promote the refractive index 
to be a complex number: $\bar{n} = n_R + i n_I$. To illustrate 
how the imaginary part of the refractive index affects the attenuation and absorption, 
let us consider an electric field $\vec{E}\,(t,\vec{x})$ in the one-dimensional case 
($\vec{x} \equiv z$)
\begin{eqnarray}
    \nonumber\vec{E}(t,z) &=& \vec{E_0}\, \exp{\left(i 2\pi \bar{n} z/\lambda - i\omega t 
    \right)} =\\
    &=&\vec{E}_0\, \exp{\left[i2\pi (n_R+in_I)z/\lambda - i\omega t \right]} = \\ 
    &=&  \vec{E}_0\, \exp{\left(-2\pi n_I z /\lambda + i2\pi n_R z/\lambda - i\omega t 
    \right)}\nonumber \equiv\\ 
    &\equiv& \vec{E}_0\underbrace{ \exp{\left(-\alpha_{atm} \cdot z/2 \right)}}_{\text{Attenuation}}
    \cdot \underbrace{\exp{\left[i \left(k z - \omega t \right)\right]}}_{\text{Propagation}}.
\end{eqnarray}
There exist a plethora of phenomena that affect the intensity loss of an electromagnetic wave propagating through the atmosphere such as heavy weather conditions like storms, foggy skies, cumulonimbus clouds forming high columns, which would represent a high loss source, rendering the satellite communication nonviable, and require precise weather forecast models; high-energy absorption phenomena like Compton scattering, photoelectric absorption and pair production (light-matter interactions), which require high energy ($\sim 60$ keV) \cite{berger2010xcom}, small wavelength ($\sim 10^{-11}$ m) electromagnetic waves, and pointing and tracking errors due to tiny fluctuations in the refractive index in small regions of the atmosphere. We may overcome the problems posed by these effects by assuming that we will work in the low-energy domain ($\lambda \sim 10^{-9}$ m) to avoid energetic absorption effects, that the line of sight between satellite and ground station will not be blocked by weather conditions, and that the pointing and tracking errors will be either negligible or corrected via correction techniques \cite{Waltenberger2026RobustPointing}. Therefore, the two main sources of intensity loss in the atmosphere we will consider are the energy absorption 
by the different atmospheric compounds and the scattering of light rays due to the
non-negligible size of the atoms and molecules. Let us split the attenuation component 
of the intensity into two parts
\begin{equation}
   I= |\vec{E_0}|^2\exp{(-\alpha_{atm}\, z )} = I_0\cdot \exp{\left[-(\alpha_{\text{Rayleigh}} + \alpha_{\text{abs}})\, z\right]}.
\end{equation}
The absorption coefficient can be sampled from NASA measures on the opacity of 
the atmosphere~\cite{NASA_atmospheric_transmittance}. To further simplify the empirical values, we approximated them by employing step functions for the values where the opacity dropped to $\sim 0 \%$, and negative Lorentzian distributions centered on the absorption wavelengths of the considered gas species within a relevant segment of the electromagnetic spectrum to model relevant atmospheric absorptions. The Rayleigh absorption coefficient can be approximated by first calculating the 
total cross-section $\sigma_{\text{Rayleigh}}$ using the following equation~\cite{Hecht2017}
\begin{equation}
    \sigma_{i,\text{Rayleigh}}(\lambda, z) =  \dfrac{8\pi}{3}\left(\dfrac{2\pi}{\lambda}\right)^4 \cdot \dfrac{n_R^2-1}{n_R^2+2}\cdot r_i^6,
\end{equation}
where $\lambda$ is lights' wavelength, $n_R$ the real part of the refractive index and $r_i$ the molecular radius of 
each gas species. Note that the term \textit{Rayleigh absorption coefficient} is not correct since Rayleigh scattering is elastic, but a \textit{tag} to emphasize that, due to the presence of this coefficient, yet another contribution to the beam intensity's loss is considered. The relation between the cross-section and the absorption coefficient is
\begin{equation}
    \alpha_{\text{Rayleigh}} = \int_0^{z_{max}} \sum_{i\in \text{gasses}}\left(\sigma_{i,\text{Rayleigh}} \cdot \rho_i(z') \, N_A /M_i \right)\,\text{d}z',
\end{equation}
where $\rho_i(z)$ are the densities of the gasses, $M_i$ the molar mass of each component, 
and $N_A$ Avogadro constant. The complete factor that modulates the amplitude of the 
electric field $\vec{E}_0$ will be called the \textit{transparency coefficient}. 
If the transparency is plotted against the wavelength we arrive at Figure~\ref{fig:transp}, 
where we can clearly see at which wavelengths the atmosphere absorbs the most.

\begin{comment}

If we want to obtain an expression for the real part of the refractive index dependent on frequency, we can compute the imaginary part using the Kramers-Kronig relations:
\begin{equation}\label{kramerskronig}
    n_I(\omega) = -\dfrac{1}{\pi}\text{PV}\int_{-\infty}^{\infty}\text{d}\omega'\dfrac{n_R(\omega')}{\omega' - \omega}\,,
\end{equation}
where $\text{PV}$ stands for Cauchy's principal value. Since we already have the real part of the refractive index of the different layers of the atmosphere, we may calculate integral \eqref{kramerskronig} to obtain the complete refractive index. In order to realistically model the absorption (and for the integral to properly converge) we will add a parameter $\    \Lambda_r$ to the denominators inside the sum in equation \eqref{sellmeier}. The imaginary part of the refractive index will then be:
    \begin{eqnarray}
\nonumber n_I(\lambda,\rho_r)=
&\dfrac{4\pi^2 m_e c^2 \varepsilon_0}{\lambda^2 e^2}\cdot
\displaystyle \sum\limits_{r=1}^{M} \rho_r
\left[
A_{1r}\lambda_{1r}^2
\dfrac{\Lambda_r}{(\lambda^2-\lambda_{1r}^2)^2+\Lambda_r^2} \right.
+\\&\left.
+A_{2r}\lambda_{2r}^2
\dfrac{\Lambda_r}{(\lambda^2 -\lambda_{2r}^2)^2+\Lambda_r^2}
        \right].
    \end{eqnarray}
\end{comment}

\subsection{Transmittance. Reflection and Transmission Coefficients}

The amount of light that travels through a medium is refereed to as transmittance 
(or transmissivity), and if we want to know how much light gets reflected or passes 
through a medium, we need to calculate the Reflection and Transmission coefficients. 
To calculate them, we resort to the Fresnel equations
\begin{equation}
    R_{\bot} \equiv\dfrac{J^r_{\bot}}{J^i_{\bot}} = \left \lvert \dfrac{n_1 \cos{\theta_i} - n_2 \cos{\theta_t}}{n_1 \cos{\theta_i} + n_2 \cos{\theta_t}}\right\rvert^2,
\end{equation}
\begin{equation}
    R_{\parallel} \equiv\dfrac{J^r_{\parallel}}{J^i_{\parallel}} = \left \lvert \dfrac{n_2 \cos{\theta_i} - n_1 \cos{\theta_t}}{n_2 \cos{\theta_i} + n_1 \cos{\theta_t}}\right\rvert^2,
\end{equation}
\begin{equation}
    T_{\bot} = 1-R_{\bot}\equiv\dfrac{J^t_{\bot}}{J^i_{\bot}}, \qquad T_{\parallel} = 1- R_{\parallel} \equiv\dfrac{J^t_{\parallel}}{J^i_{\parallel}},
\end{equation}
where $J^{r/t}_{\bot/\parallel}$ is the reflected/transmitted energy flux of the 
perpendicular/parallel component of the wave, defined as the dot product of the 
Poynting vector $\vec{S}$ with the direction of propagation of the wave. The closer 
the transmission coefficients are to the value of 1.0, the more intensity passes 
through a medium.  In this work, the computation of the transmission coefficients will enable us to analyze what percentage of photons passes through the atmosphere under any conditions at a given altitude $z$ and wavelength $\lambda$ by resorting to equation~\eqref{eq:index}. The next task 
is addressing light polarization, which is a key part in QKD protocols, and the 
laser's behavior. 
%\begin{figure}[hbtp]
%    \centering
%    \includegraphics[width=0.99\linewidth]{Figures/transmission_vs_altitude.png}
%    \caption{Perpendicular and parallel (cumulative) transmission coefficients against altitude %for a wavelength of $\lambda=1000$ nm and for an incident angle of $45.0^\text{o}$.}
%    \label{fig:tarnsm}
%\end{figure}

\subsection{QKD protocol E91 and depolarization}

There are a plethora of quantum applications and protocols that could be tested in order to validate the model and to create a framework with which to evaluate viability. Our simulator had already implemented a working E91~\cite{E91_Ekert} protocol. This presented an opportunity to compare against real world data taken from satellite entanglement distribution, as exposed in the paper by Yin \textit{et al.}~\cite{yin2020entanglement}. Specifically, this work employs the Micius satellite as a middle point to distribute entanglement between two ground stations 1,120 kilometers apart. Even though in the aforementioned work BBM92~\cite{bennett1992quantum} is used, owing to the fact that both protocols perform entanglement distribution to generate the key, we consider that the comparison should yield similar metrics in terms of QBER and Secret Key Rate (SKR). Because of this shared core principle and provided that our experiments were designed to match the exact parameters of this reference setup, with as much precision as Netsquid~\cite{netsquid} (our simulation tool) allows, we argue that simulation results in this context should adjust to the empirical data and as such, should validate our model.

In regards to the specific Netsquid requirements, we used the depolarizing channel since depolarization is a major impediment to E91, the reason lying in the nature of the protocol. When measuring with their respective bases (of which only two coincide), depolarization produces a loss of correlation or anti-correlation that tampers with expected QBER values, preventing adequate identification of eavesdroppers in the channel. For this depolarizing channel model Netsquid requires a depolarization rate ($R_{\text{pol}}$) that introduces a probability of depolarization whose expression is deduced from the Lindblad master equation. The complete Lindblad Master equation reads ($\hbar=1$)
\begin{equation}
\label{eq:Limblad}
    \dot{\rho} = - i\left[H,\rho\right] +\dfrac{R_{\text{pol}}}{4}\sum_{i} L_i\rho L^\dagger_i - \dfrac{1}{2}\left\{L^\dagger_i L_i\,, \rho \right\},
\end{equation}
where $R_{\text{pol}}$ is the depolarization rate, $\rho$ is the density matrix, $L_i$ are the jump operators, $H$ is the time-independent closed system's Hamiltonian and $\left[\,, \right]$, $\left\{\,, \right\}$ stand for the commutator and anticommutator, respectively. By choosing Pauli matrices as our jump operators $L_i = \sigma_i$, the interaction term on \eqref{eq:Limblad} becomes
\begin{equation}\label{eq:Limblad_sigma}
    \dot{\rho} = \dfrac{R_{\text{pol}}}{4}\sum_{i} \sigma_i\rho\sigma_i - \dfrac{1}{2}\sigma_i^2\rho -\dfrac{1}{2}\rho\sigma_i^2, = \dfrac{R_{\text{pol}}}{4}\sum_{i} \sigma_i\rho\sigma_i -\rho,
\end{equation}
since $\sigma_i^\dagger = \sigma_i$ and $\sigma_i^2=\mathbb{I}$ where $\mathbb{I}$ represents the $2\times 2$ identity matrix, the anticommutator becomes $\left\{\mathbb{I}\,, \rho\right\} \equiv 2\rho$. Now, we employ the relation $2\mathbb{I}-\rho = \sum \sigma_i \rho \sigma_i$, and~\eqref{eq:Limblad_sigma} becomes
\begin{equation}\label{eq:limbad_diff}
    \dot{\rho} = -i\left[H,\rho\right] +\dfrac{R_{\text{pol}}}{2} (\mathbb{I}-2\rho).
\end{equation}
Equation \eqref{eq:limbad_diff} is a first order, ordinary, %linear,
differential matrix equation which can be solved by employing the following change of variables to reduce it to the standard von Neumann's form for closed systems
\begin{eqnarray}
    \rho(t) \rightarrow \dfrac{\mathbb{I}}{2}+ \bar{\rho}(t) \cdot\exp\left(-R_{\text{pol}}\cdot t \right),
\end{eqnarray}
and, by applying the properties of commutators on the new variables $\left[H,\rho\right] = \exp\left(-R_{\text{pol}}\cdot t \right) \cdot \left[H,\bar{\rho}\right]$, we arrive at

\begin{equation}
    \dot{\bar{\rho}} \equiv -i\left[H,\bar{\rho}\right]
\end{equation}
%\begin{eqnarray}
%    -2\int_{\rho_0}^{\rho} (\mathbb{I}-2\rho')^{-1}\text{d}\rho ' &=& \ln\left(\dfrac{\mathbb{I}-2\rho}{\mathbb{I}-2\rho_0}\right) =  -R_{\text{pol}}(t-t_0) \Rightarrow\nonumber \\
%    \Rightarrow\mathbb{I}-2\rho &=& \left(\mathbb{I}-2\rho_0\right)\exp\left[-R_{\text{pol}}(t-t_0)\right]\nonumber,
%\end{eqnarray}
which can be easily solved on the Schrödinger picture as
\begin{equation}
    \bar{\rho}(t) = U(t)\,\bar{\rho}(0)\,U^\dagger(t),
\end{equation}
where $U(t) = \exp\left( -iHt\right)$ stands for the time-evolution operator. Finally, to solve for $\rho(t)$, we must apply the inverse change of variables noting that $\bar{\rho}(0) = \rho(0)- \mathbb{I}/2$, yielding the final expression
\begin{equation}
    \rho(t) = \dfrac{\mathbb{I}}{2} + U(t)\left(\rho(0) - \dfrac{\mathbb{I}}{2}\right) U^\dagger(t)\cdot \exp\left(-R_{\text{pol}}\cdot t \right).
\end{equation}
Therefore, the depolarization will trigger under the influence of a decaying exponential which depends on $R_{\text{pol}}$. This exponential is the method Netsquid uses to evaluate the depolarization probability. Notwithstanding this effort, we shall as well justify which depolarization probability we chose. This depolarization probability can be deducted by the polarization extinction ratio (PER), this figure is defined as the ratio of powers of different polarizations. It characterizes the degree of polarization of the devices. To match the PER that should be expected in the Micius experiment, we referred to the work authored by Wu \textit{et al.}~\cite{Wu:20} analyzing the optical elements pertaining to the Micius experiment. Defining an expected QBER due to depolarization that essentially matches the depolarization probability $\left[\text{QBER}_{\text{pol}} \sim 1- \exp{\left(-R_{\text{pol}}\cdot t\right)}\right]$, allows us to, from a reference PER, obtain the depolarization rate. Essentially $\text{QBER}_{\text{pol}}\approx1/\text{PER}$, this simplification is explained in their work and relies on the fact that the minimum required PER is high valued with respect to 1, and that deviations between intended and real polarization values are negligible with the equipment that is used. Depending on the delay the beam experiences the actual probability and thus, depolarization rate varies. The variations of the value of $R_{\text{pol}}$ near the visual range of the electromagnetic spectrum is small enough to consider $R_{\text{pol}}$ as a constant and, %this is, $R_{\text{pol}}^{Max} - R_{\text{pol}}^{Min} \sim 0.1$ \textcolor{blue}{Sergio revisa estos valores pls}; and so, 
as a general figure and due to time constraints on the code implementation, we took $R_{\text{pol}}=8\text{ Hz}$, which corresponds to higher depolarization probabilities than those experienced in the travel distances in the orbit points in between ground stations (corresponding to 5.2 Hz or 2\% depolarization probability approximately) which, in our opinion, balances the optimistic predictions the model asserts and allows us to treat transmissions within an expected conservative depolarization range throughout all simulations. Additionally, since the value of $\text{QBER}_{\text{pol}}$ exponentially depends on $R_{\text{pol}}$, small changes on the value of the depolarization rate will not trigger significant changes on the QBER values. We consider that this choice balances the optimistic predictions the model asserts and allows us to treat transmissions within an expected conservative depolarization range throughout all simulations.

\subsection{Theoretical model for the laser: Gaussian Beams}
The easiest way to model a real light beam is using the Gaussian Beam description. 
A Gaussian beam is described by an electric (magnetic) field whose amplitude envelope 
in the transverse plane is given by a Gaussian curve. One of the most important 
consequences of this profile is the fact that the cross-section of the laser does not 
increase in a linear fashion, instead the cross-section decreases until a minimum section 
called \textit{waist} $w_0$ and then starts increasing again. The equations for the 
Gaussian beam traveling in the $z$ direction are~\cite{Siegman1986}
\begin{eqnarray}
    \nonumber\displaystyle \vec{E}(r,z) =\dfrac{\vec{E}_0 \cdot w_0}{w(z)} \exp \left(
    \frac{-r^2}{w^2(z)} 
    -\frac{\alpha_{atm} z}{2}\right) \cdot \\
    \cdot \exp \left[i\bigl(kz - \omega t + \frac{kr^2}{2R(z)} - \xi(z)\bigr) \right],
\end{eqnarray}
where $w(z) = w_0 (1 + (z/z_0)^2)^{1/2}$, $z_0 = \pi n_R w_0^2/\lambda$,
$R(z) = z (1 + (z_0/z)^2)]$, and $\xi(z) = \arctan{(z/z_0)}$,
and, the intensity of the beam is given by
\begin{eqnarray*}
    I(r,z) &\equiv\dfrac{\vec{E}\cdot\vec{E^*}}{2\eta_w}= I_0\left[\dfrac{w_0}{w(z)}\right]^2 \exp{(-\alpha_{\text{atm}} \, z -2r^2/w^2(z))},
\end{eqnarray*}
$\eta_w$ is the wave impedance of the medium in which the beam is propagating.

\subsection{Optical link power model}

We turn now our attention to the satellites used in the communication. The usual 
optical link can be modeled by an equation, a product of efficiencies and losses that 
relates transmitted power to the received power, this is a modification of the Friis 
equation for the link budget, the specific modifications applied for the case of 
optical communications is extracted from Hemmati's book on Deep Space Optical 
Communication~\cite{Hemmati2006}, as for all expressions in this section
\begin{equation}
    P_S = P_T\,\dfrac{G_T\,G_R\,\lambda^2}{16\pi^2 r^2}\cdot \eta_T \, \eta_a \, \eta_r \cdot 
    L_{T_p} \, L_{\text{atm}} \, L_{\text{pol}} \, L_{R_p},
\end{equation}

 where $\eta_{(\cdot)}$ stands for the efficiency of the transmitter's optics, 
 aperture illumination and receiver's collector respectively, $A_{(\cdot)}$ is 
 the aperture are for transmission and reception and $L_{(\cdot)}$ stands for 
 the different sources of power loss. The transmitter and receiver gains can respectively be defined as $G_T=4\pi A_T/\lambda^2$ and $G_R=4\pi A_R/\lambda^2$; the terms $A_T,A_R$ are the apertures of both the transmitter and receiver. 

\begin{comment}

\begin{equation}
P_S=P_T\cdot\eta_T\cdot\eta_a\cdot\frac{4\pi A_T}{\lambda_T^2}\cdot L_{Tp}\cdot L_{atm} \cdot L_{pol} \cdot L_{Rp}\cdot\frac{A_R}{4\pi\cdot r^2}\eta_r ,
\end{equation}

where:
\begin{itemize}
    \item $\eta_t$: Efficiency of the transmitter's optics.
    \item $\eta_A$: Aperture illumination efficiency.
    \item $\eta_r$: Efficiency of the receiver's collector.
    \item $A_T$: Aperture Area for transmission.
    \item $A_R$: Aperture Area for reception.
    \item $L_{Tp}$: Transmitter pointing losses.
    \item $L_{atm}$: Atmospheric absorption losses.
    \item $L_{pol}$: Polarization losses.
    \item $L_{Rp}$: Receiver pointing losses.
\end{itemize}
\end{comment}

In terms of space communications, augmenting the transmitted power could be a
good idea to increase the received power, but it would be restricted to the power
budget requirements. Another idea could be increasing the aperture area, which would 
reduce the beam's width but, in return, increases the probability and magnitude of 
the PAT error.

Since we are working with the gaussian beam formalism,  the expression for the 
aperture illumination efficiency is
\begin{equation}
    \eta_A= \frac{2S}{\alpha^2}\left[\exp{(-\alpha^2 \gamma^2)}-\exp{(-\alpha^2)}\right],
\end{equation}
where $\gamma$ refers to the obscuration ratio, $S$ is the RMS optical path 
difference (for smooth optics, its value is 28\% of the peak to valley value), 
$\alpha \equiv A_d/2w(z)$ where $A_d$ corresponds to the aperture diameter 
and $w(z)$ is the beam's radius.

\begin{comment}
    \begin{equation}
    \alpha=\frac{A_d}{2w(z)},
\end{equation}

\begin{equation}
    S=e^{(\frac{-2\pi \sigma}{\lambda})^2}.
\end{equation}

In this context:

\begin{itemize}
    \item $\gamma$: This is the obscuration ratio.
    \item $\sigma$: RMS optical path difference. For smooth optics, is usually 28\% of the peak to valley value.
    \item $A_d$: Aperture diameter.
    \item $B_w$ Gaussian beam width.
\end{itemize}

\end{comment}

\subsection{Background interference sources}

Most of the interference can be boiled down to diffused light from the atmosphere 
($P_{\text{diffuse}}$), or sourced from the planets ($P_{\text{planetary}}$), 
scattered light in the optics ($P_{\text{stray}}$). These can be expressed as
\begin{subequations}
    \begin{align}
    &P_{\text{diffuse}} = L_\lambda(\theta) \cdot A_R\cdot\Omega\,\Delta \lambda \cdot \eta_R, \\
    &P_{\text{planet}} = H(\lambda)A \,\Omega\Delta \lambda \cdot \eta_R, \\
    &P_{\text{stray}} = I_\lambda \cdot\Omega\Delta \lambda \,\, A\cdot \eta_R \cdot T(\lambda) \cdot BSDF(\theta),
    \end{align}
\end{subequations}
where $H(\lambda)$ is the spectral irradiance of the background, $L_\lambda$ refers 
to the sky irradiance, $I_\lambda$ is the exo-atmospheric solar constant 
($0.074 \, \text{W/cm}^2 \cdot \mu \text{m}$), BSDF($\theta$) is the Directional 
Scatter Distribution, which follows a power law distribution as a function of $\theta$ 
and $T(\theta)$ being the atmospheric absorption.

\begin{comment}

\begin{itemize}
    \item $H(\lambda)$: Spectral Irradiance of the background.
    \item $L_\lambda$: Sky irradiance (function of the wavelength and solar illumination geometry).
    \item $I_\lambda$: Exo atmospheric solar constant (0.074W/cm$^2\mu m$).
    \item BSDF($\theta$): Directional Scatter Distribution. Follows a power law distribution as a function of $\theta$.
    \item $T(\theta)$: Atmospheric absorption.
\end{itemize}

\end{comment}

\section{Specifications and Technical Details}
\label{sec:specifications}

To correctly address design goals and measure accurate results that validate 
the model, it is important to create a reference framework. Testing specific cases 
and results would only be accurate if the model tried to replicate the conditions 
of a certain set of experimental results. Of course, a certain abstraction level 
is needed to reduce or neglect some variables or take into account others. In terms 
of satellite quantum communications we have to take into account some limitations when 
designing a framework. First, technical standards do not meet established legal 
or commercial regulations, neither is there a specific standard that is commonly and 
internationally accepted. This is unlike other telecom services where specific power, 
precision and wavelengths would exist. Second, each country with existing quantum 
communication projects may have  different providers, goals and requirements, so 
two completely different sets of parameters can be completely valid because they aim 
to fulfill different goals or different conditions. Thirdly, there are some 
countries that do have such a significant presence in the space sector, that their 
technologies  and standards can get imposed as a side effect, with no guarantee that other 
future projects will adopt it. Finally, the actual constraints of the different subsystems 
of a space system vastly differ in the case of a ground station and the satellite by 
itself. This asymmetry  is key to correctly understand the nature of the communication links. 

%What we have aimed for is creating, first, a fundamental set of parameters that are relevant to the model or that directly or indirectly affect one or more parameters inside the model. And second, a range of coherent values for these parameters that are coherent when combined with one another. The goal is, in this line of thought, to be able to test approximate scenarios for relevant experiments and, inside the established range, compare the obtained results. 

\subsection{Optical Space Segment Specifications}

Space segment specifications represent the maximum constraint that the link reaches, 
as the power budget in satellites and deployed systems are strict. Apertures and 
efficiencies in the satellite are actually the most important parameters to take into 
account because the modules have to fulfill a given set of requirements, which 
conditions shapes, mass, volume, materials, and many other variables. As such, one 
cannot simply assume an aperture of an arbitrary number of meters, because for that 
arbitrarily large telescope to be orbiting there would need to exist appropriate 
equipment to actually deploy it. Looking at QKD projects, in the case of the Micius
satellite a 300-mm-aperture Cassegrain telescope was used and a 5 $\mu$W laser was 
used for tracking, the QKD link was operating at 850 nm. The diffraction loss was 
estimated to be 22 dB for a link of 1200 km, the tracking precision was 1.2 micro 
radians~\cite{Liao2017}. The Micius satellite is actually a relatively large 
(over 500 kilograms) satellite in comparison to other initiatives. In contrast, 
the QUBE-II~\cite{QUBEII}, which is based on the CubeL mission 
(2021), for CubeISL, and the precursor mission QUBE-I (2024) 
which follows the CUBESAT standard~\cite{CubeSatInfo}, it represents a radical shift 
in paradigm compared to Micius. Firstly, CubeSats operate in different 
standarised categories that are internationally respected, from U1 ($10 \times 10 \times 10$ 
cm for 1 to 1.33 kg) to U6 or U12 ($20 \times 20 \times 30$ cm for 20 to 50 kg). The 
difference between categories is remarkable, and actually, U6 U12 CubeSats are in fact 
deployed for commercial use with specialized modules. The QUBE-I operates at 850 nm 
and has an aperture of 50 mm. Larger satellites can host more equipment and are relatively 
less constrained in terms of power. As a result, the non-standardized satellites 
are not only more technologically advanced and thus less restricted, but this develops 
in such an advantage over CubeSats, that large satellites dedicated to QKD can be 
deployed to geostationary orbits.

% This initiates an interesting discussion. In one hand, large companies and 
% governments can afford million to a hundred million dollar projects that develop 
% medium to large satellites without standardization, this is an obstacle for open
% sourced models and software because details and parameters do not only follow a 
% different set of constraints, but in many cases are not entirely public. But this 
% has other implications, 

This is a great advantage in terms of precision but it is not attainable for 
all latitudes. This is because after a certain lower bound elevation angle, 
the distance that the photons need to travel over the atmosphere is so large 
that the transmitivity vanishes. This "atmospheric wall" effect 
supposes a great disadvantage for GEO satellites because they inherently need 
to orbit in the Earth's plane. This means that for many regions of the planet
direct line of sight (LOS) is heavily impeded for GEO quantum communication. 
For example, a 30 mm aperture transmitter/receiver 
in a CubeSat supposed to operate in GEO is not realistic because 
it would not be able to compensate the "atmospheric wall" effect. This could 
be easily overcome by running the model and measuring the received power. 
For LEO and MEO satellites, using apertures like 300 mm or more can be unrealistic, 
specially for LEO. While for bigger apertures, better yields are expected, 
the satellites that are usually deployed in these conditions are restricted in 
terms of space, shape, and volume. In other words, when working 
with a CubeSat of a given category, then we know that the aperture does not only 
need to be proportional to the size of the CubeSat specification, but that it needs
to fit its size. For satellites developed and operated by government actors,
these constrains are less of an issue. For instance, the Hispasat QKD-GEO satellite 
project aims for GEO QKD with the requirements seen in Table~\ref{hispasatTable}.

\begin{table}[t]
\centering
\small % Reducing font size slightly to fit the width
\begin{tabularx}{\columnwidth}{>{\columncolor{hispasatblue}\color{white}\bfseries}l X}
\hline
\rowcolor{hispasatblue} \multicolumn{2}{c}{\color{white}\bfseries SYSTEM PARAMETERS} \\ \hline
Orbit & GEO \\
Position & $30^\circ$ W -- $61^\circ$ W \\
Power &  258 W (TBC) \\
Mass & 212 kg (TBC) \\
QKD Tech. & Discrete Variable \\
Protocol & BB84 + Decoy \\
Link Budget & 47 dB (TBC) \\
Wavelength & 1550 nm \\
Telescope aperture & 50 cm \\
Key Rate & 30--500 bps \\
Key Volume & 1.5--3 Gb \\
Pointing Accuracy & 1 $\mu$rad  \\
Services & Commercial \\ \hline
\end{tabularx}
\caption{Extracted from \cite{abad2024hispasat}, from the QKD-GEO Hispasat.}
\label{hispasatTable}
\end{table}

%A valid approach to validate or implement simulations is to try to reproduce the expected results from such experiments or taking the requirements as a basis to build a framework. In summary, apertures in satellites range from the tens of millimeters to 50 cm or more, but being heavily reliant on the size and orbit of the satellite. Precision ranges from the tens of micro radians to less than or equal to one micro radian and it implies the existence of an ATP module. A realistic simulation would take into consideration the consumption of the modules and subsystems of the space segment in terms of power budget. This power budget can be approximated if not defined by the example of other experiments.

%\subsection{Ground Segment Specifications}

%In contrast to the Space Segment the Ground Segment equipment is more permissive and has less parameters to take into consideration. The previously mentioned Hispasat QKD-GEO has specifically, an aperture of 1.8 meters, a pointing accuracy of 1 $\mu$rad and of course, works in the same wavelength.

\begin{figure*}[t]
    \centering
    \includegraphics[width=0.45\linewidth]{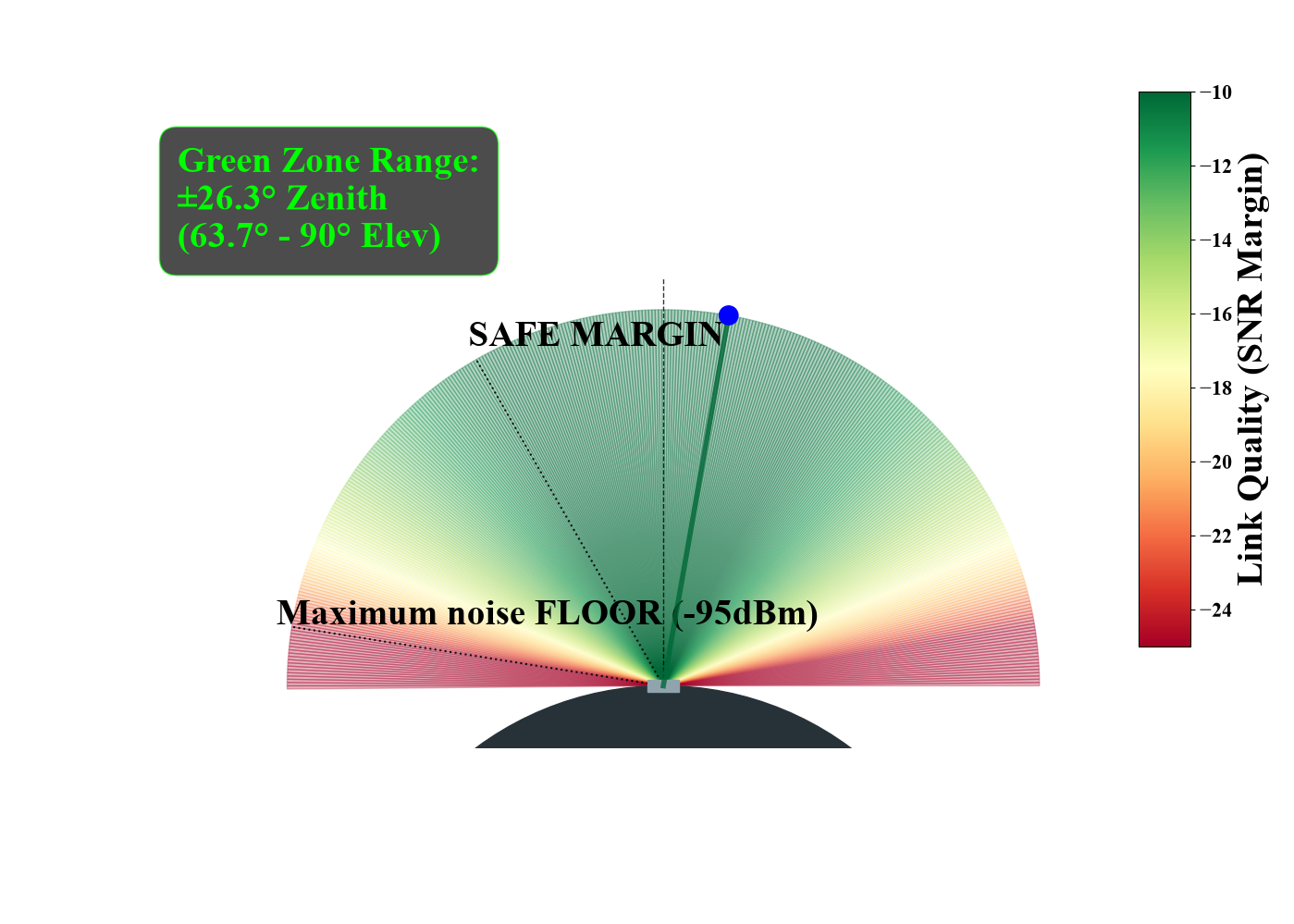}
    \includegraphics[width=0.45\linewidth]{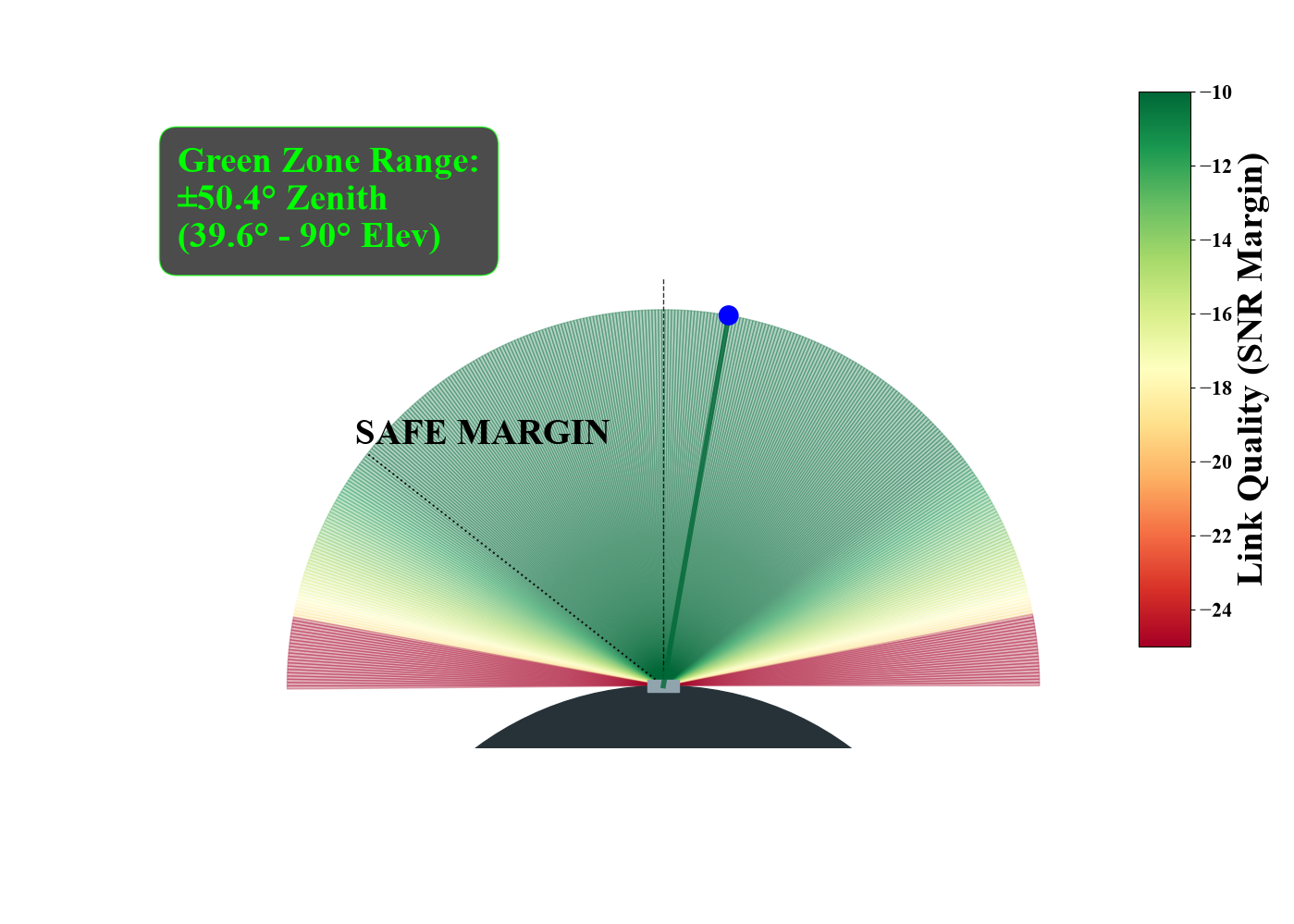}
    \caption{Sketch of the perceived power for wavelegnths $\lambda = 1550$ nm (left) and $\lambda=800$ nm (right) received in the satellite due to all incorporated effects taken into account in the model, greener zones represent bearable losses in the corresponding angles. The precision of the equipment involves a mean angular error of approximately 5 micro radians working with a 1550 nm laser.}
    \label{fig:1550nmLaserLosses}
\end{figure*}

\subsection{Terrestrial Network Integration}

Integration with terrestrial networks needs to balance the use of the different 
wavelengths that are usable in the context of QKD. As the wavelength increases 
in value, the required precision to operate also increases. This means that the greater 
the wavelength (lower frequencies) the higher the precision the equipment must attain. 
This is caused by the dispersion of the laser. Recalling the expression of the Gaussian 
Beam, dispersion depends on the value of $z_0$ which constrains the waist of the beam 
and how it propagates onward. To better understand this, 
Figure~\ref{fig:1550nmLaserLosses} depicts the two main approaches to quantum 
communications. Basically, smaller wavelengths have a greater divergence angle, 
needing more precise optical equipment, Equipment becomes precise after the 
implementation of advanced ATP systems (Advanced Tracking and Pointing System), 
to levels below the micro radian.  

% TODO: Add reference to SPADS
Equipment produced to work in the 800 nm (although not this exact wavelength 
but approximated) is far more efficient and cheap. The 1550 nm wavelength is however, 
much better suited to work in existing fiber optic systems. Thus, it is a matter of
reconciliation between network domains. In summary, free space quantum communications 
works in the 800 nm mark because of the efficiency and low power consumption of the 
emitters compensating the higher dispersion rate with technology such as ATP systems, 
this tradeoff is beneficial but has the side effect of isolating free space networks 
from ground networks operating with fiber optics in terms of wavelength. This interfacing 
problem is not an issue. Ground Stations can exchange information with both ground 
and space network (at the cost of slight increase in complexity).

Our model reflects this tradeoff between the wavelength and the precision. For each 
actual scenario the designer would have to take into account the expected values of 
dispersion of the beam and the exact wavelength the equipment is working on. 

%Existing
%technology achieves precisions that can be effectively abstracted to a precision level 
%ranging from few micro radians to the hundreds of nano radians. The lack of specificity 
%is once again, due to the great variations produced by different technologies, 
%configurations, operating conditions, and other characteristics.  

\section{Model Results}
\label{sec:results}
\subsection{Atmospheric test for Micius and Hispasat missions}
Based on the previously discussed conditions for the different projects we decided 
to test the model results by simulating mainly two specific cases, the Hispasat and 
the Micius satellite. On one hand, the Micius satellite, with abundant information of 
technical details and results is the ideal case to reproduce for our model. Comparing 
the results of the Micius experiment and the result of our model provides a unique 
opportunity for validation, as previously discussed. The Hispasat satellite does not have such an enormous 
base of scientific literature and measurements, but poses a similar initiative yet
at the same time radically distinct approach (due to the changes in altitude, 
wavelength and receiver and transmitter specifications). Based on this premise,  
we have produced a series of figures as the output of the simulation based on the 
developed model, for which we have taken into account both the downlink and the 
uplink for each of the cases (Micius and Hispasat).

% Conditions of the experiment 
For the Hispasat case a wavelength of 1550 nm was used as indicated by the 
requirements of the project, in the case of the Micius case, a 810 nm wavelength, 
in both cases the same ground station was employed (1 m in aperture). Despite 
both GS having different specifications for this parameter we wanted to use a 1 m 
aperture as a compromise between both cases, by not using bigger apertures we aimed 
to produce more conservative results. The satellites' apertures are sampled from the original works 
regarding the Hispasat and Micius missions~\cite{abad2024hispasat, RevModPhys.94.035001}, 
whose values are $50$ cm and $30$ cm, respectively.

\begin{figure*}
    \centering
    \includegraphics[width=0.45\textwidth]{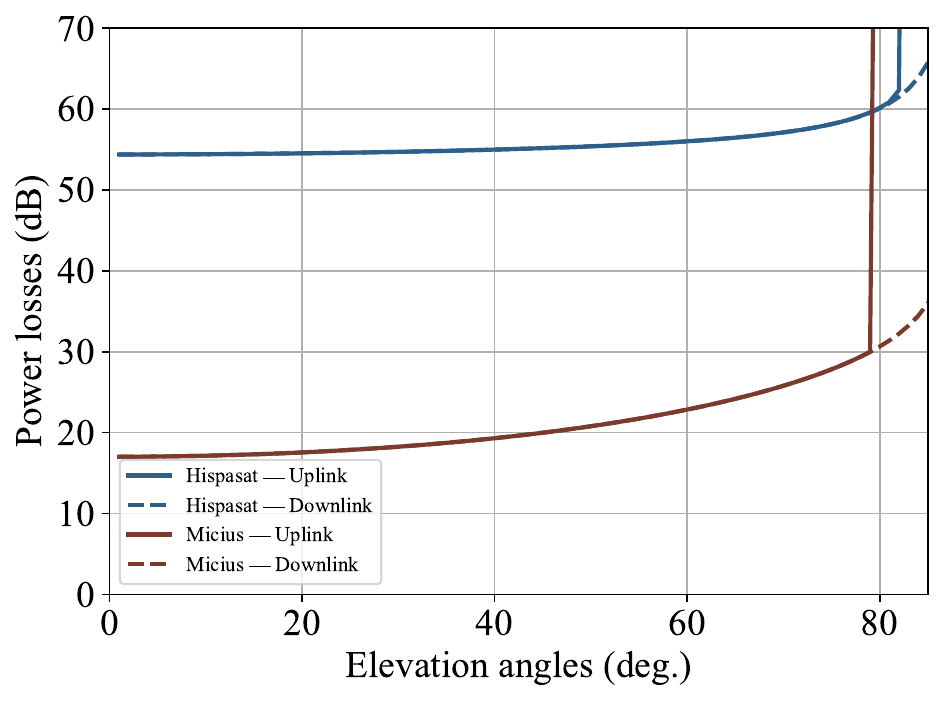}
    \includegraphics[width=0.45\textwidth]{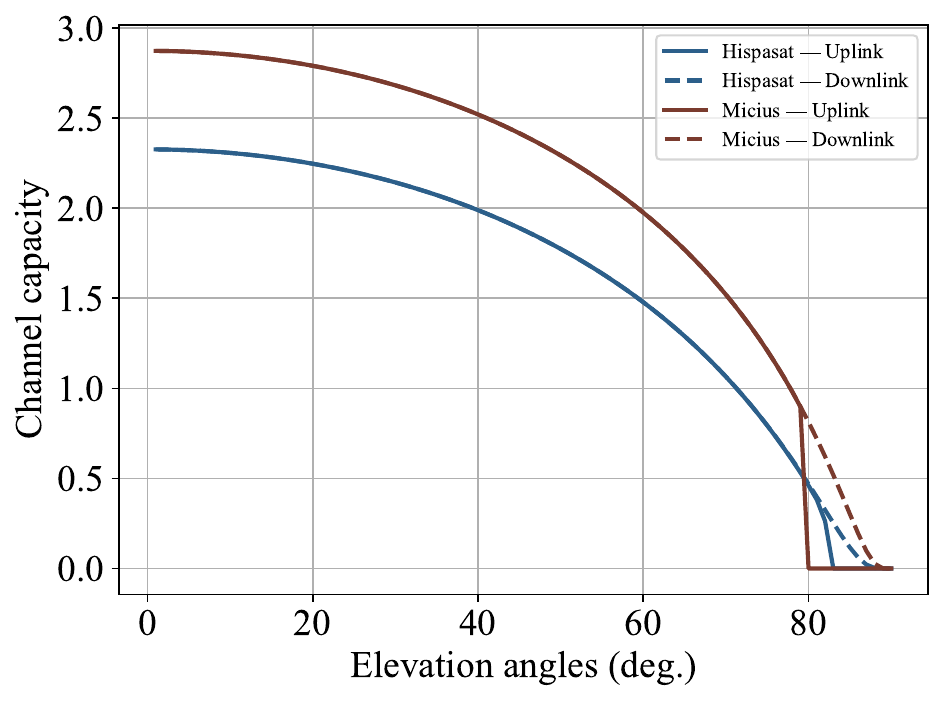}
    \caption{Power losses in dB and channel capacity as functions of the elevation angle.}
    \label{fig:power_losses_vs_angles_def}
\end{figure*}

Figure~\ref{fig:power_losses_vs_angles_def} shows the previously mentioned atmospheric 
wall effect, by which after a certain elevation angle in the uplink transmission, the
initial segment of the communication experiments huge losses as the beams approach the 
horizon. Downlink losses are not nearly as severe for these extremes angles because 
relative to the total distance of the link, the final segment (the atmosphere in that case) 
is minuscule. Still, as the angles reach 90 degrees, the losses rise asymptotically.

\begin{figure*}
    \centering
    \includegraphics[width=0.45\linewidth]{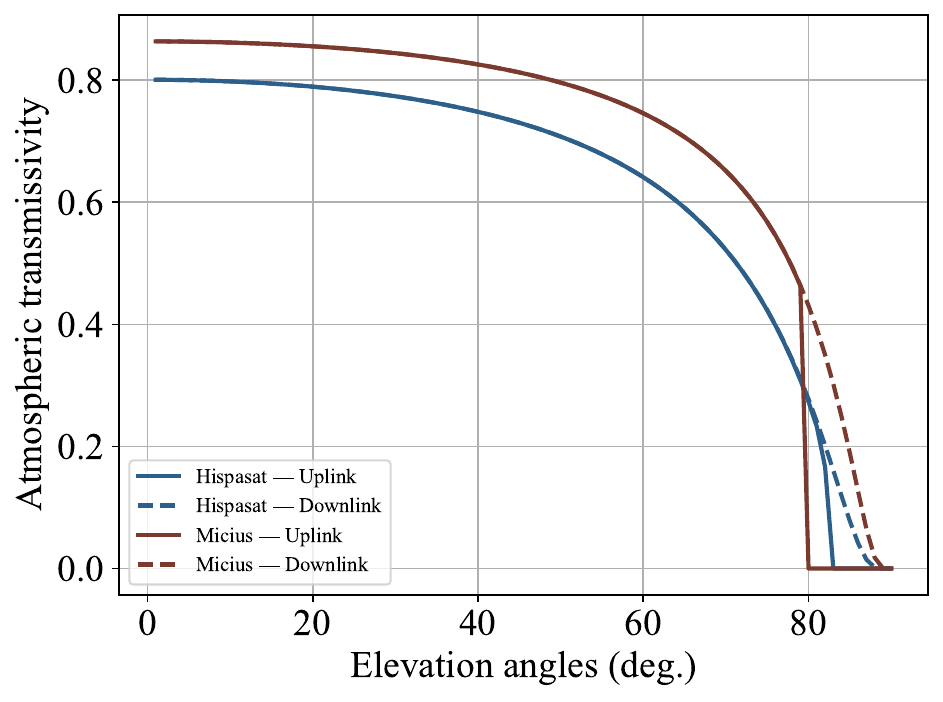}
    \includegraphics[width=0.45\linewidth]{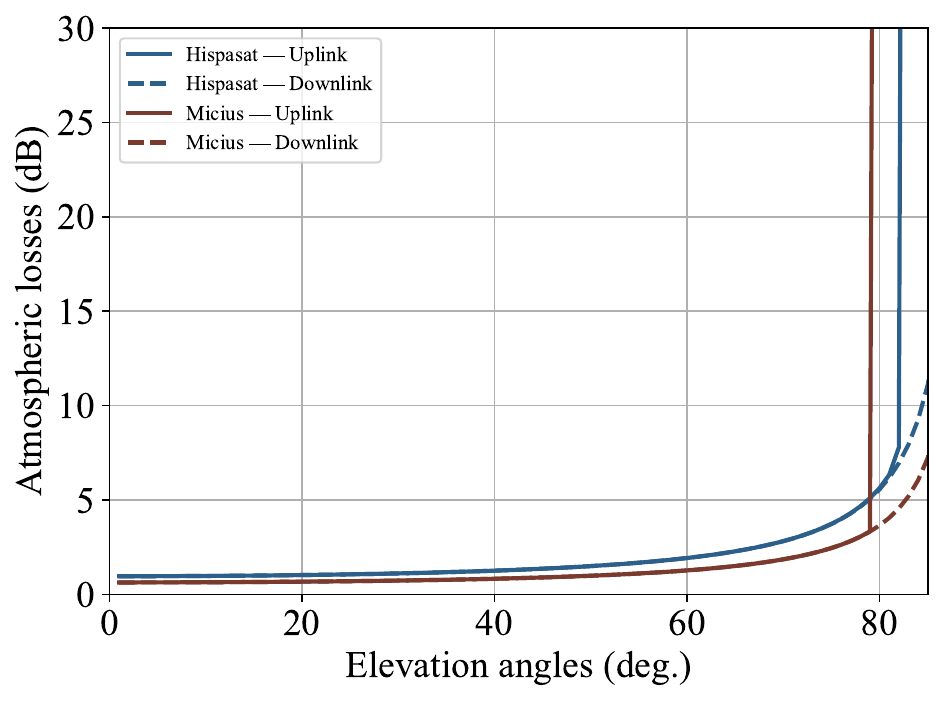}
    \caption{Transmissivity $T$ and atmospheric losses (dB) 
    as a function of the elevation angle.}
    \label{fig:transmissivity_vs_angles_def}
    %\caption{Atmospheric loss (dB) vs. elevation angle.}
    %\label{fig:atm_losses_vs_angles_def}
    %\caption{Power received in dBm versus the elevation angles.}
    %\label{fig:power_received_vs_angles_def}
\end{figure*}

Beyond basic power losses we can separate this result into other variables to 
see exactly why the model behaves like this. Figure~\ref{fig:transmissivity_vs_angles_def} 
shows how tranmissivity evolves for both Micius and Hispasat, the wall effect is more 
remarkable here for the Micius case while Hispasat manages to hold better transmissivity 
values across the collapsing region which supports the idea that the requirements and the 
concept of the mission conveyed, Hispasat's QKD-GEO aimed for GEO quantum communications, 
if we take into account the latitude of a country such as Spain (which is much closer 
to the equator compared to other Northern European ones) and the fact that the 
transmissivity holds for longer in this case we can deduce that such a communication 
scheme is valid , even though Hispasat shows a lower transmissivity in general to that 
of the Micius satellite.  In contrast, downlink does not show much variation for the 
same reasons as the loss of power did.

In the case of the losses that the atmosphere shows, in 
Figure~\ref{fig:transmissivity_vs_angles_def} (right), these are a perfect negative image 
to that of the power loss. Specially so because in our model's simulations we have not 
taken into account statistical losses for other phenomena not relative to the atmosphere,
however, the difference in atmospheric losses between the cases are not as remarkable 
as in the power losses. This could be explained by the fact that in both cases we are 
operating at wavelengths at which the atmosphere behaves very similarly. The wall effect 
is also visible here and the uplink/downlink asymmetry is also present.

%As we have seen these graphs present the same information but on different levels,  allowing to contemplate the nuances of the link for different cases. The last of the  figures takes the same concept, in Figure~\ref{fig:transmissivity_vs_angles_def} (right) 
%we  show how much power gets to the receiver taking a reference of 1 Watt emitted power.

%The previously mentioned figure can be especially helpful when designing the link budgets, Hispasat accounted for a link budget of 47 dB, for a 1 W power emission in the laser, the resulting value exceeds even the most conservative margins. This shows that we could afford to emit far less power so as to tighten the link budget, traditionally, a 6 dB margin is enough to take into account most adverse phenomena, although for space communications this margin could be exceedingly optimistic. 

These results can be contrasted to the predictions of the Micius experiments. 
In terms of diffraction, approximately $22$ dB were calculated, and absorption in 
atmosphere losses ranging from $3$ dB to $8$ dB. Our predictions for atmospheric losses 
in the elevation angles that Micius operated in (10º to 15º) our model predicts around 
$-3$ dB, that is coherent, specially taking into consideration that in our model there 
are other phenomena that were not taken into account, many 
contributions that despite being small individually, may amount to compensate with 
this offset, which despite existing, is within a reasonable margin of error. Total 
losses were calculated to be in the range $-31$ dB to $-39$ dB approximately. Our model's  
prediction stays at $-29$ dB. Again, taking into account that the model lacks certain 
phenomena, and that the conditions for the simulation were similar but not identical 
so as to be compared with the hypothetical Hispasat case, we think that the predictions 
are in fact reasonable. Figure~\ref{fig:nnew_power_losses} specifically targets the 
range that the Micius satellite works in, with the exact parameters that the Micius has. 
The difference is now much higher (the model predicts a $-27$ dB to $-30$ dB loss), and 
this is due to this lack of modeled phenomena. Results still approximate to the correct 
range, so while concrete and precise values are not possible with the actual state of 
the model, we are certain that with more refined additions to the code these could be 
achieved. In terms of the transmissivity according to the atmosphere losses, 
with absorption losses of $-3$ dB to $-8$ dB, that implies a transmissivity in the range 
of $0.5$ to $0.16$, our prediction stays at the $0.5$ to $0.46$ range approximately, so 
in terms of transmissivity we conclude that the predictions are close to
experimental results.

%\begin{figure*}[t]
%    \centering
%    \includegraphics[width=0.45\textwidth]{Figures/nnew_power_losses_vs_angles_def.png}
    
%    \caption{Power loss vs. elevation angles for the Micius case.}
%    \label{fig:nnew_power_losses}
    %\caption{Channel capacity against elevation angles.}
    %\label{fig:capacity}
%\end{figure*}

\subsection{Link capacity based on model results}

Power budgets and transmissivity are a helpful approach towards the exploration 
of viable quantum communication link, but the analysis beyond viability pertains 
more expressions and relationships beyond this. For instance, capacity establishes 
a limit to the size of the possible key established by quantum key distribution 
protocols (QKD). For the capacity we can use the Pirandola–Laurenza–Ottaviani–Banchi 
(PLOB) bound~\cite{Pirandola2019} $C(T)=-\log_2(1-T)$, where $T$ is the total 
channel transmissivity (Figure~\ref{fig:power_losses_vs_angles_def}, right). In our case this means 
we will be dealing with the  transmissivity of the atmospheric link, which is 
generated by the simulation model. 
In terms of longer chains of repeaters, such as multi-satellite communication networks 
this can simply be reduced to the minimum capacity link, meaning that low yield 
links will produce a bottleneck in the network reducing the upper bound of the key 
significantly. This, together with our observations of the model results in the Uplink 
capacity being the bottleneck (so long as there are not worse links once in space, 
which is unlikely).

\subsection{Satellite QKD comparison with Micius}

To assess the practical viability of E91-based quantum key distribution (QKD) under realistic atmospheric conditions, we performed a series of numerical simulations to reproduce the scenario entanglement distribution scenario between the two ground stations and satellite involved in the aforementioned Micius experiment using our atmospheric channel model discussed on Section~\ref{sec:model}.

Simulations were carried out for elevation angles ranging from ($\varepsilon_1 \in [120^\circ, 10^\circ]$) measured from one of the ground stations, with a satellite altitude of $500$ Km and a wavelength of $\lambda= 810$ nm, consistent with the LEO configuration employed in the Micius satellite experiments~\cite{RevModPhys.94.035001}. For each elevation angle we computed the the expected channel transmittance, the resulting quantum bit error rate (QBER) and the sifted key rate following the standard E91 security analysis.
Table~\ref{tab:skr_qber_elevation} presents the corresponding Secure Key Rate in bits per second and QBER of said angles.

The QBER obtained from our simulations remains below the $10\%$ threshold, allowing for sufficient preemptive margin for eavesdropper detection. In contrast with real Micius data, our results predict lower QBER (at an average of 1\% error rate), but for the reference real values that were measured to have that level of QBER the SKR relies close below the values, thus being more pessimistic. We performed 200 repetitions for each point. 

\begin{table}[t]
\centering
\small % Matches reference table font size
\begin{tabularx}{\columnwidth}{>{\columncolor{hispasatblue}\color{white}\bfseries}c *{4}{>{\centering\arraybackslash}X}}
\hline
\rowcolor{hispasatblue} \multicolumn{5}{c}{\color{white}\bfseries SKR AND QBER ANALYSIS} \\ \hline
\rowcolor{hispasatblue} \color{white}\bfseries Elevation ($^\circ$) & \color{white}\bfseries SKR & \color{white}\bfseries SKR Std.Dev. & \color{white}\bfseries QBER & \color{white}\bfseries QBER Std.Dev. \\ \hline
117 & 0.30 & 0.10 & 0.010756 & 0.004940 \\
85  & 0.35 & 0.14 & 0.007855 & 0.004941 \\
54  & 0.20 & 0.06 & 0.008672 & 0.005117 \\
36  & 0.24 & 0.09 & 0.007482 & 0.004962 \\
24  & 0.32 & 0.14 & 0.008358 & 0.004359 \\
17  & 0.28 & 0.09 & 0.009408 & 0.005044 \\
\hline
\end{tabularx}
\caption{SKR and QBER Analysis by Elevation Angle.}
\label{tab:skr_qber_elevation}
\end{table}

The disagreements between the experimental results and our simulated values are small in absolute value and originate from our assumption that the satellite remains in the same position at the instant of time of the entanglement distribution for a given transmission of a key block (which is needed in order to use Netsquid, which does not incorporate dynamic systems), and consequently pointing and tracking errors are not realistic, which is an unavoidable limitation of static simulators. A noteworthy observation is that the channel shows more SKR when the satellite is between ground stations, in terms of transmissivity both uplink and downlink have a similar value in this region, amounting to a higher probability that both photons arrive at their destination. However it is also true that because of the slower fall in transmissivity as a function of the elevation angle in the downlink channel, that also has the effect of increasing SKR in positions that are better for the uplink, as that does not lower the downlink's transmissivity as much. Overall, these results support the conclusion that E91-based QKD over a satellite link is numerically consistent with the performance demonstrated by the Micius experiment, and validate our atmospheric model as a reliable and open tool for predicting the feasibility of future satellite-based QKD missions.

\subsection{Starlink exploratory assessment}

\begin{figure*}[t]
    \centering
    \includegraphics[width=0.45\textwidth]{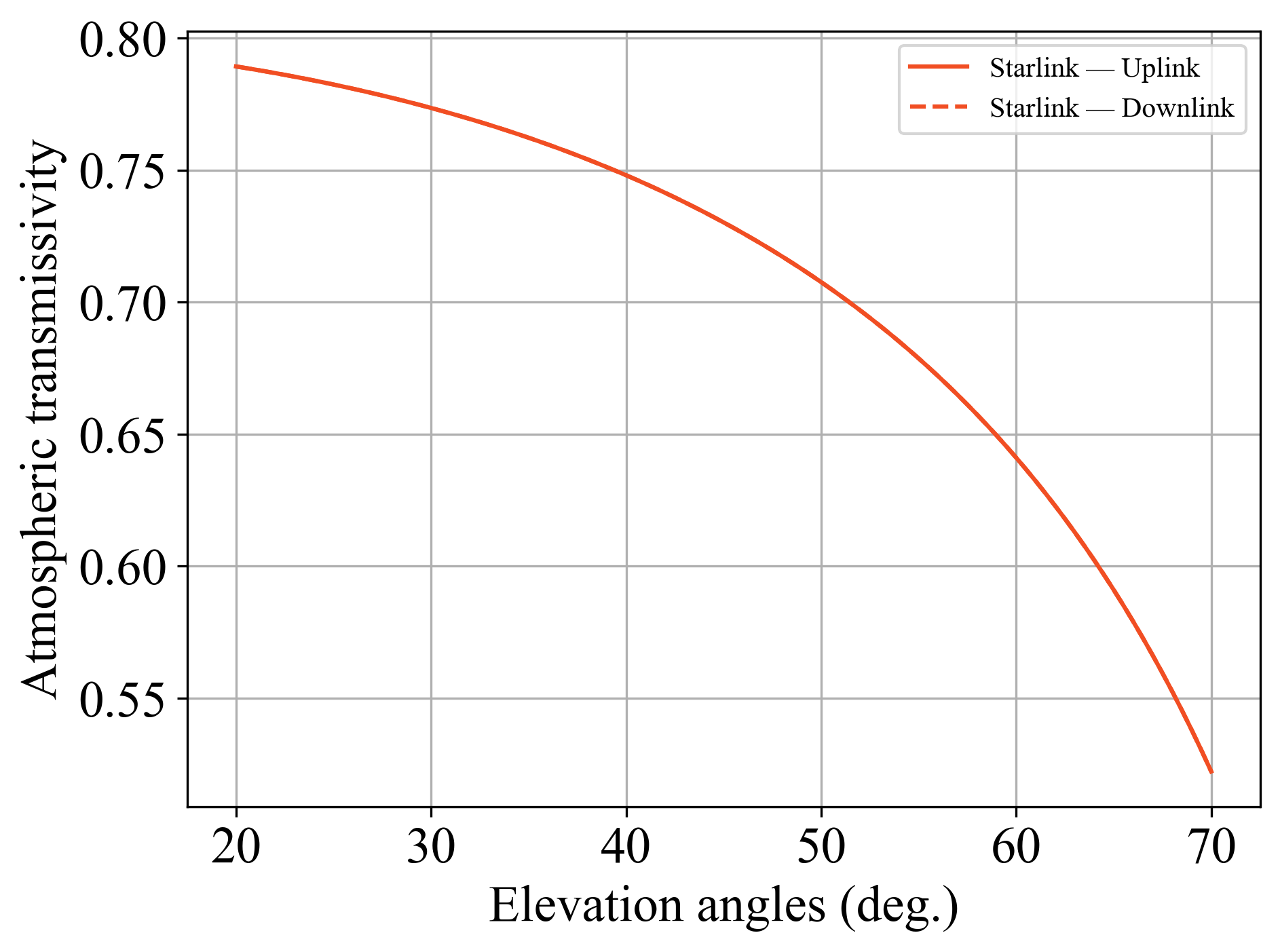}
    \includegraphics[width=0.45\textwidth]{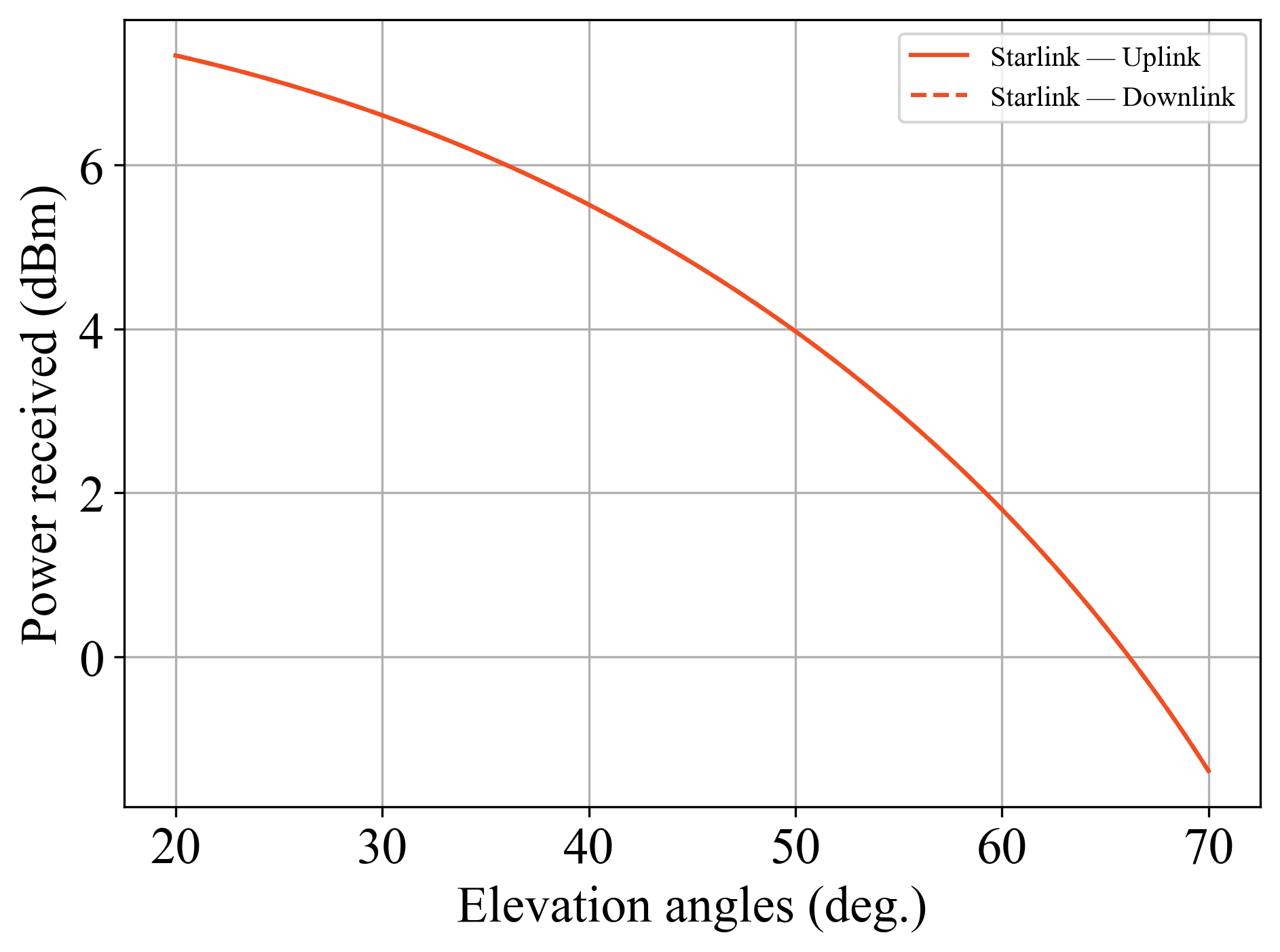}
    \caption{Atmospheric transmissivity and power vs. elevation angles for the Starlink satellites.}
    \label{fig:S_trans}
\end{figure*}

The Starlink constellation comprises LEO satellites at altitudes ranging from
340 km to 570 km, with orbital inclinations designed to provide near-global coverage. 
Equipped with the results provided by the atmospheric channel model, we will study how 
feasible quantum communications employing the Starlink constellation are. We must have in mind for the purposes 
of simulation, that we considered similar satellite specifications as the Micius mission 
within significant changes on altitude (550 km from Earth's surface), wavelength
($\lambda = 1550 \, \text{nm}$) and elevation angles ($\theta \in [20^\circ, 70^\circ]$) this due to the lack of existing reference infrastructure for quantum communications in the Starlink constellation, which means: there is no existing quantum communication infrastructure, and existing infrastructure for classical communication cannot be used as a reference.
We performed simulations to obtain the atmospheric transmissivity and power balance 
to be able to plot the computed values against the different elevation angles $\theta$ 
to study the optimal window of operation, similar to previous cases.
The simulated results can be observed in Figure~\ref{fig:S_trans} where we can clearly 
notice that, if such a Starlink network is developed, it would need to 
respect this technological restrictions imposed by the specifications of the Micius 
ground station and the computed received power. In other words, mission requirements 
for a quantum communication network deployment would need to take into account how 
the received power decreases with the inclination angles.

The restriction in terms of throughput for these networks and megaconstellations could be severely limited depending on the particular protocol, but also in terms of the endpoints of the communication. Such low throughput (inferior to 1 bit/s) as seen in the validation phase of this work, corresponds to entanglement distribution networks where satellites and intermediaries between ground stations. As such, the effective probability of such a communication to be successful can be roughly described as the probability of success of both photons to arrive correctly at their corresponding ground stations. If one is lost during transmission, the pair is effectively lost. Notwithstanding these effects for ground to space quantum communications, it is true that performing ground to space key generation, or in-space quantum applications, would allow for far greater throughput, moreover, constellations and megaconstellations would allow for near perfect channels where only pointing and tracking errors worsen the rate of transmission. With highly populated megaconstellations these errors can be greatly reduced as the distance between satellites decreases. Providing a promising framework for quantum communications in space.

%\begin{comment}

\section{Conclusions}
\label{sec:conclusions}

In this work, we have developed a theoretical framework to predict results regarding transmissivity, atmospheric and total power losses for satellite quantum communication scenarios and compared against the Micius satellite for validation and Hispasat's QKD-GEO as a way to analyze the viability of the established requirements. We used publicly available information to program our model. Our results match the observations from Micius and Hispasat, demonstrating that the model adjusts to experimental outcomes.

\begin{comment}
QBER and SKR calculation remain outside of the scope of such a tool, but in fact could be calculated using other simulation tools that do require a value of transmissivity or losses. Future additions to the program's source code can contribute to this effort by adding more complex phenomena, which will make the model more precise and more precise. As other quantum communication projects arise and are studied, the results of said experiments can contribute towards its validation and troubleshooting.
\end{comment}

\section*{Acknowledgments}
This work was supported in part by MCIU/AEI/10.13039/501100011033 (DISCOVERY Project) under Grant PID2023-148716OB-C31 and by the Galician Regional Government under Project ED431B 2024/41 (GPC).

%This research project was made possible through the access granted by the Galician Supercomputing Center (CESGA) to its supercomputing infrastructure. The supercomputer FinisTerrae III and its permanent data storage system have been funded by the NextGeneration EU 2021 Recovery, Transformation and Resilience Plan, ICT2021-006904, and also from the Pluriregional Operational Programme of Spain 2014-2020 of the European Regional Development Fund (ERDF), ICTS-2019-02-CESGA-3, and from the State Programme for the Promotion of Scientific and Technical Research of Excellence of the State Plan for Scientific and Technical Research and Innovation 2013-2016 State subprogramme for scientific and technical infrastructures and equipment of ERDF, CESG15-DE-3114.

\bibliographystyle{IEEEtran}
\bibliography{References}{}
\end{document}